\documentclass[fleqn,usenatbib]{mnras}

\usepackage{newtxtext,newtxmath}
\usepackage[T1]{fontenc}
\usepackage{multirow}

\DeclareRobustCommand{\VAN}[3]{#2}
\let\VANthebibliography\thebibliography
\def\thebibliography{\DeclareRobustCommand{\VAN}[3]{##3}\VANthebibliography}

\usepackage{graphicx}	
\usepackage{amsmath}	

\title[Search of radiation from SGR1935+2154]{About the search for pulsed radiation from the magnetar SGR1935+2154 at the LPA LPI}

\author[E. A. Brylyakova]
{E. A. Brylyakova, $^{1}$
{
S. A. Tyul'bashev$^{1}$ \thanks{E-mail: serg@prao.ru}
}
\\
$^{1}$ P.N. Lebedev Physical Institute of the Russian Academy of Sciences, Astro Space Center, Pushchino Radio Astronomy Observatory,\\
Radiotelescopnaya 1a, Moscow reg., Pushchino, 142290, Russia\\ 
}

\date{October 13, 2023}
\pubyear{ , 2023}

\begin{document}
\label{firstpage}
\pagerange{\pageref{firstpage}--\pageref{lastpage}}
\maketitle

\begin{abstract}
The magnetar (J1935+2154) is known as a source of soft gamma radiation. For the first time, radio emission from SGR1935+2154 in the form of a hyperflare was detected at a frequency of 1.25 GHz on the FAST radio telescope in May 2020. The magnetar enters the survey conducted on the Large Phased Array (LPA) radio telescope at a frequency of 111 MHz. A check of the previously published (Fedorova and Rodin, Bulletin of the Lebedev Physics Institute, 2021) pulse from the magnetar SGR1935+2154 was carried out.
The data received on the LPA is recorded in parallel in two modes having low and high frequency-time resolution: 6 channels with a channel width of 415 kHz and a time resolution of $\Delta$t = 100 ms; 32 channels with a channel width of 78 kHz and a time resolution of $\Delta$t = 12.5 ms. The original search was carried out using data with low time-frequency resolution. The search for dispersed signals in the meter wavelength range is difficult, compared with the search in the decimeter range, due to scattering proportional to the fourth power of frequency and dispersion smearing of the pulse in frequency channels proportional to the second power of frequency. In order to collect a broadened pulse signal and obtain the best value of the signal-to-noise ratio (S/N), the search was carried out using an algorithm based on the convolution of multichannel data with a scattered pulse pattern. The shape of the template corresponds to the shape of a scattered pulse with a dispersion measure ($DM$) of 375 pc/cm${^3}$. For repeated verification, the same data was used in which the pulse from the magnetar was detected. An additional check of the published pulse was also carried out using data with a higher frequency-time resolution. 
Since the dispersion smearing in the frequency channel in the 32-channel data is 5 times less than in the 6-channel data, an increase approximately 2 times in the S/N pulse could be expected. Pulse radiation with S/N>4 having a pulse peak shift depending on $DM$  from SGR1935+2154 was not detected in either 32-channel or 6-channel data.

\end{abstract}



\section{INTRODUCTION}

In 2007, fast radio bursts (FRBs) were discovered by \citeauthor{Lorimer2007} (\citeyear{Lorimer2007}). The nature of FRB remains unclear and there are number of hypotheses about their origins \citeauthor{Petroff2022} (\citeyear{Petroff2022}). FRBs are powerful single radio pulses with durations ranging from fractions of a millisecond to several milliseconds, with a dispersion measure ($DM$) suggesting an extraterrestrial origin of the bursts. $DM$ is a characteristic of the medium that determines the different arrival times of pulses at different frequencies due to the frequency-dependent plasma refractive index.

By now, a number of extragalactic FRBs have been detected in the decimeter range, including repeating ones \citeauthor{Spitler2016} (\citeyear{Spitler2016}); \citeauthor{Andersen2019} (\citeyear{Andersen2019}); \citeauthor{Fonseca2020} (\citeyear{Fonseca2020}). One hypothesis regarding the nature of FRBs is magnetar giant flares \citeauthor{Popov2010} (\citeyear{Popov2010}). The first detection of a magnetar giant flare in the radio range was recorded in April 2020 when a FRB was detected from the Galactic magnetar SGR1935+2154, previously known as a source of soft gamma-ray emissions \citeauthor{Zhang2020} (\citeyear{Zhang2020}). The registered FRB burst was preceded by a series of soft gamma-ray bursts observed by the Swift satellite \citeauthor{An2020} (\citeyear{An2020}).

In 2021, a study was published \citeauthor{Fedorova2021} (\citeyear{Fedorova2021}) in which a fast radio burst was detected at a frequency of 111 MHz using the Large Phased Array (LPA) of the Lebedev Physical Institute (LPI) of the Russian Academy of Sciences with a $DM$ of 320 pc/cm${^3}$, which the authors associated with the source SGR1935+2154. In this study, we conducted a verification of the detected pulse using the same data as used by Fedorova and Rodin, and by repeating their proposed search methodology. It should be noted that previous checks of the fast radio bursts detected by the authors had already been performed \citeauthor{Brylyakova2023} (\citeyear{Brylyakova2023}), which showed the absence of reliable signals in the specified region.

\section{VERIFICATION OF THE FOUND FRB}

The search for FRBs was conducted in the data of daily 24/7 monitoring carried out at the LPA LPI radio telescope in 96 spatial beams. This monitoring is part of the PUMPS survey (Pushchino Multibeam Pulsar Search, \citeauthor{Tyul'bashev2022} (\citeyear{Tyul'bashev2022}) and started in August 2014. The LPA LPI is a meridian-type radio telescope consisting of 16,384 dipoles. The beamwidth is approximately $0.5^{\circ} \times 1^{\circ}$, and the source passes through the meridian in about 3.5 minutes at half-power. The central observation frequency is 110.3 MHz, with a bandwidth of 2.5 MHz.

The search for FRBs, according to the work of \citeauthor{Fedorova2021} (\citeyear{Fedorova2021}), and the references therein, was performed as follows: a segment corresponding to the source's passage through the LPA beam pattern was selected from the hourly recording; data were convolved with a template matching the presumed signal shape from SGR1935+2154, according to the scattering model by \citeauthor{Kuzmin2007} (\citeyear{Kuzmin2007}); visual inspection for pulses was carried out for the period from September 1, 2019, to February 12, 2021, resulting in the detection of a pulse on September 2, 2020.

Monitoring data are recorded in two modes simultaneously, with low and high frequency-time resolutions: 6 channels with a channel width of 415 kHz and a time resolution of $\Delta$t = 100 ms, and 32 channels with a channel width of 78 kHz and a time resolution of $\Delta$t = 12.5 ms. Data with low frequency-time resolution are used in the ''Space Weather'' program (\citeauthor{Shishov2016}, \citeyear{Shishov2016}) and for monitoring observation quality. Data with high frequency-time resolution are used for pulsar and transient searches. \citeauthor{Fedorova2021} (\citeyear{Fedorova2021}) used low frequency-time resolution data for the FRB search. Using the same data, we attempted to detect the burst on September 2, 2020.

According to a previously published work of \citeauthor{Deneva2016} (\citeyear{Deneva2016}), the radio telescope's sensitivity for single dispersed pulses can be calculated using the formula:

\begin{equation}
    {S_{\min} = \frac{W_{\text{obs}}}{W_{\text{int}}} \times \frac{kT_{\text{sys}}}{A_{\text{eff}} \sqrt{N_{\text{pol}} \Delta\nu W_{\text{obs}}}}},
	\label{eq:eq_1}
\end{equation}
Here, $W_{\text{obs}}$ and $W_{\text{int}}$ are the observed and intrinsic pulse widths, \(k\) is the Boltzmann constant, $T_{\text{sys}}$ is the system temperature (for estimates, 1000 K), $A_{\text{eff}}$ is the effective area of the LPA (for estimates, 47,000 sq.m.), $N_{\text{pol}}$ is the number of polarizations (one), and $\Delta \nu$ is the total observation bandwidth (2.5 MHz). In turn, the observed pulse width is determined by the point sampling time ($\Delta t$), dispersion smearing in the frequency channel ($t_{\text{ch}}$), pulse broadening due to scattering ($t_{\text{kuz}}$), and intrinsic pulse width ($W_{\text{int}}$):

\begin{equation}
    {W_{\text{obs}} = \sqrt{\Delta t^2 + t_{\text{ch}}^2 + t_{\text{kuz}}^2 + W_{\text{int}}^2}},
	\label{eq:eq_2}
\end{equation}

From equations (1, 2), it is evident that sensitivity depends on the apparent width of the observed pulse. The lower the impact of scattering and dispersion smearing in the frequency channel, the higher the sensitivity of the observations. Lower observation frequencies are more affected by scattering and dispersion smearing. This impact is particularly significant in the meter range. $DM$ increases the pulse width due to dispersion smearing in the frequency channels, as $t_{\text{ch}} \propto \text{$DM$}  \times (\Delta\nu / \nu^3)$, and scattering increases the pulse width, like $t_{\text{kuz}} \propto (\text{$DM$}/100)^{2.2}$.

The wider the channel width, the more pronounced the dispersion smearing, which means that for narrow pulses, where the impact of dispersion smearing and scattering is low, the sensitivity will be higher. It is also evident that for the case of a narrow pulse with a duration less than \(\Delta t\), the sensitivity will be higher for data with high frequency-time resolution.

Using equation (1), theoretical sensitivity was estimated for data with high and low frequency-time resolution obtained in the LPA monitoring, depending on $DM$. Figures 1-2 depict the dependence of sensitivity and pulse broadening for 6- and 32-channel data. From Figure 1, it can be observed that for $DM$ $\approx$ 300 pc/cm${^3}$, the sensitivity in 32-channel data is significantly higher than in 6-channel data. At $DM$ $\approx$ 600 pc/cm${^3}$, the difference in sensitivity between 32- and 6-channel data becomes negligible. This is because, for $DM$ > 330 pc/cm${^3}$, the scattering of the pulse becomes so significant (see equation (1)) that the pulse width is more determined by scattering. Therefore, in the end, sensitivities are equalized. However, at $DM$ = 320 pc/cm${^3}$ (the dispersion measure of the magnetar), the sensitivity in 32-channel data is approximately twice as high as in 6-channel data.

\begin{figure}
\begin{center}
	\includegraphics[width=\columnwidth]{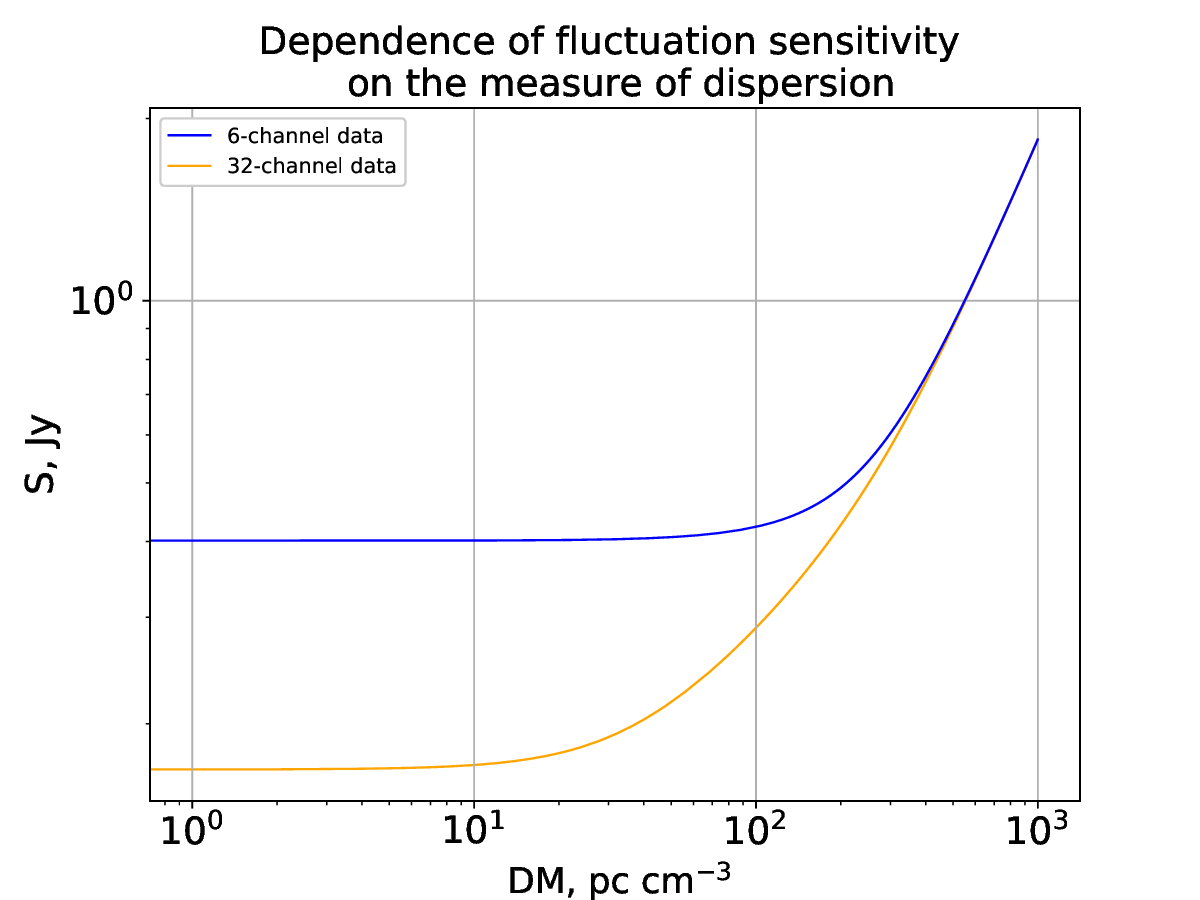}
    \caption{The sensitivity of the LPA LPI radio telescope, calculated using formula (1), for 6- and 32-channel data.}
    \label{fig:Ffig1}
\end{center}
\end{figure}

\begin{figure*}
\begin{center}
	\includegraphics[width=\textwidth]{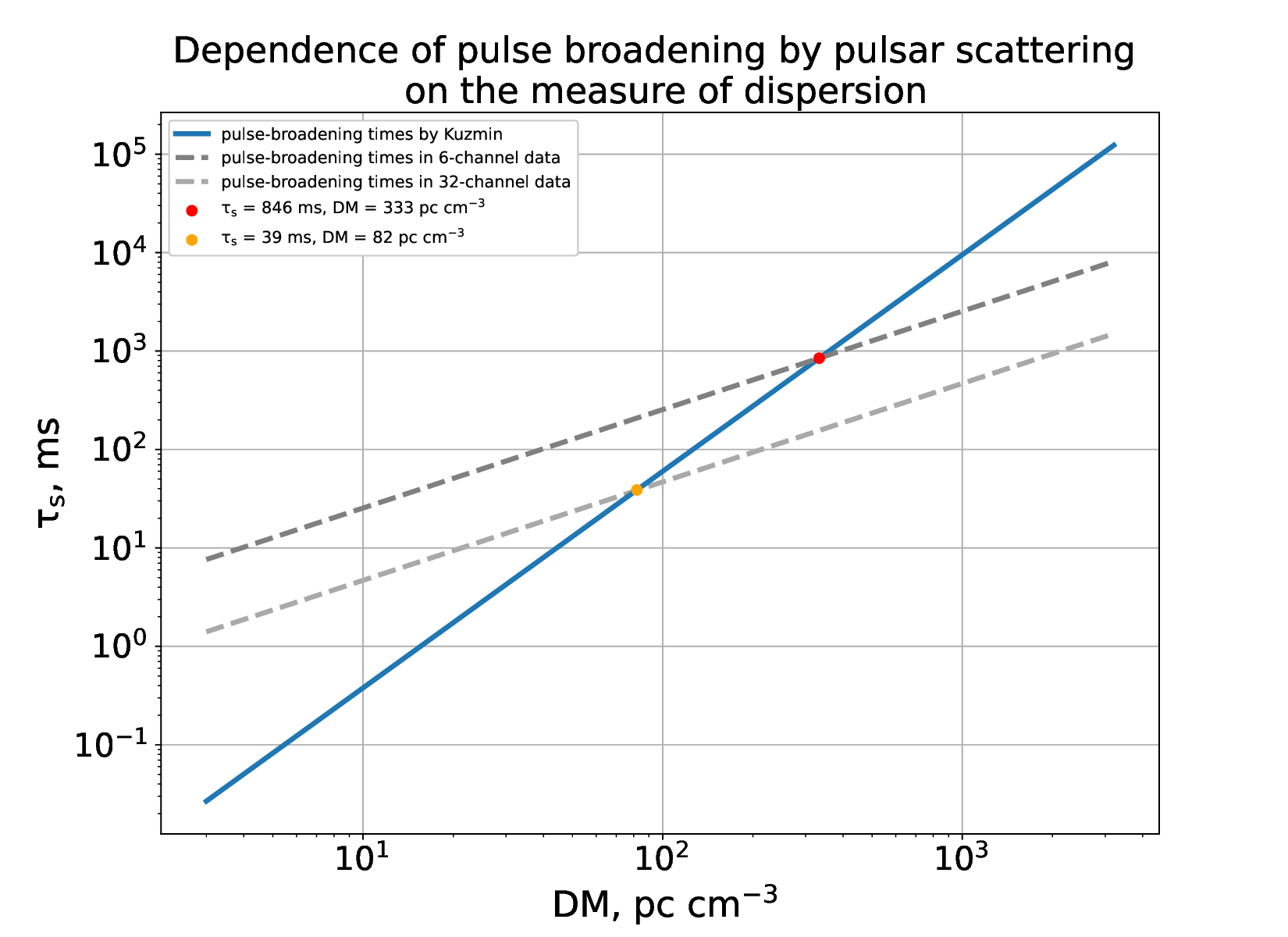}
    \caption{The pulse broadening due to dispersion smearing (shown by dark and light gray dashes for 6- and 32-channel data) and due to scattering according to the Kuzmin model (\citeauthor{Kuzmin2007}, \citeyear{Kuzmin2007}).}
    \label{fig:fig2}
\end{center}
\end{figure*}

\begin{figure*}
    \centering
    \begin{minipage}{0.49\textwidth} 
        \includegraphics[width=\linewidth]{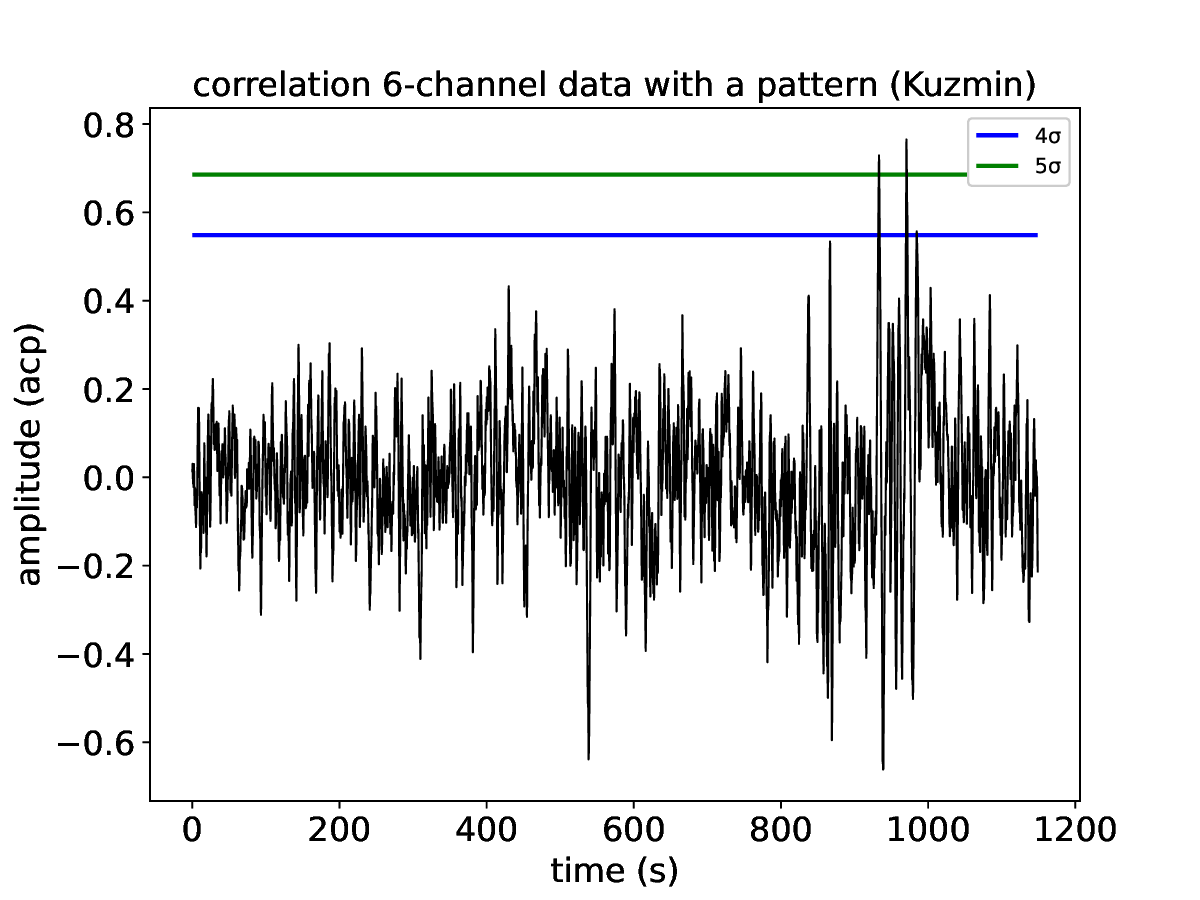}
    \end{minipage}
    \hfill 
    \begin{minipage}{0.49\textwidth} 
        \includegraphics[width=\linewidth]{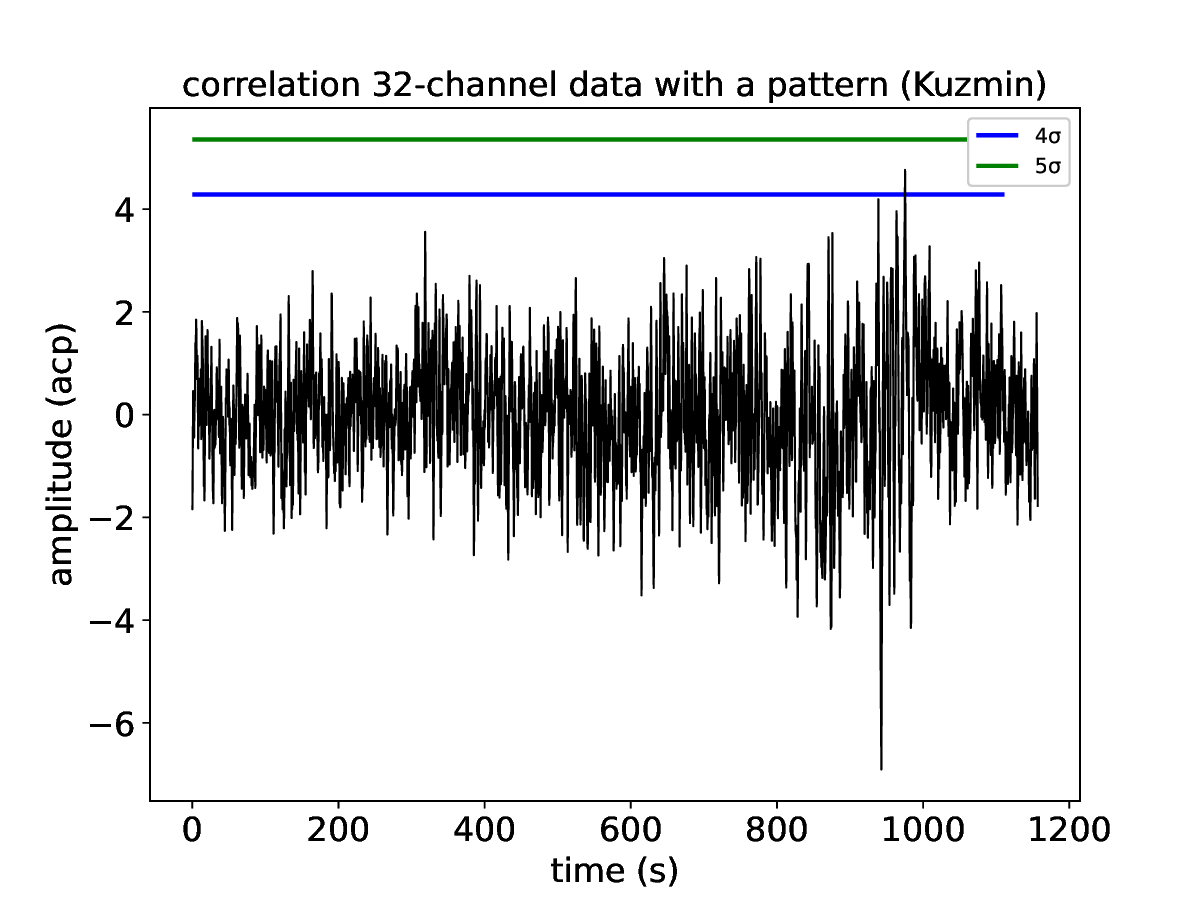}
    \end{minipage}
    
    \caption{20 minutes of processed data after convolution with the Kuzmin template: a - 6-channel data, b - 32-channel data. The center of the figure corresponds to the magnetar's coordinates.}
    \label{fig:рисунок1_2}
\end{figure*}

Repeating the methodology of \citeauthor{Fedorova2021} (\citeyear{Fedorova2021}), we processed both 6-channel and 32-channel data, assuming that the S/N would be higher in the 32-channel data. To examine the vicinity of the source, a 20-minute recording was chosen for processing. Unfortunately, we were unable to detect a dispersed signal above 4 sigma of the noise level in the direction of the magnetar. Figure 3 shows the processed recording using the authors' proposed methodology. In the processed 6-channel data, there is a signal at a level of $ 5 \sigma{_n}$, but the distance to this signal is 8 time minutes, which exceeds the beamwidth of the LPA LPI.
  
\section{DISCUSSION OF THE RESULTS}

The search for FRBs in the meter range presents a complex technical challenge. Firstly, the greater the $DM$, the more significant the dispersion smearing in the frequency channels. To avoid sensitivity loss due to dispersion smearing, Western colleagues working in the meter wavelength range on the LOFAR radio telescope use narrow frequency channels (\citeauthor{Pastor-Marazuela2021}, \citeyear{Pastor-Marazuela2021}). Secondly, using arbitrarily narrow frequency channels still cannot compensate for signal broadening due to scattering. Scattering broadens the signal and degrades sensitivity. When processing monitoring data from the LPA LPI, we encounter both of these factors. Since the majority of known FRBs have $DM>300-400$~pc/cm$^3$, it is scattering that will determine the final sensitivity in their search. \citeauthor{Fedorova2021} (\citeyear{Fedorova2021}) believe that convolution with the template of the scattered pulse allows them to collect the signal without a loss in the S/N. However, collecting the signal without losses is only possible in cases where the duration of the pulse is longer than the characteristic scattering time. It is known about the magnetar J1935+2154 that since 2020, multiple radio flashes have been registered from it. According to published data (\citeauthor{Andersen2019}, \citeyear{Andersen2019}; \citeauthor{Bochenek2020}, \citeyear{Bochenek2020}; \citeauthor{Kirsten2021}, \citeyear{Kirsten2021}), the typical duration of the pulses is fractions of a millisecond. Thus, the pulses found by other researchers for this magnetar are inherently narrower than the scattering time at the observing frequency of the LPA and 10-100 times shorter than the sampling in the LPA data, with high and low frequency-time resolution, respectively. There is no reason to assume that this magnetar has pulses wider than a second, which is precisely the pulse width needed to achieve the same S/N for processed LPA data. Therefore, if the pulse indeed exists, then in high frequency-time resolution data, it should have a S/N twice as high comparable with low frequency-time resolution data. This is not observed in Figure 3. Therefore, we consider the detection to be false. Earlier, we also attempted to detect signals from the FRBs published by the authors, and it was also unsuccessful (\citeauthor{Brylyakova2023}, \citeyear{Brylyakova2023}). Out of the three reprocessed candidates, we obtained a similar record for only one case, but the S/N was half of the value published in the paper \citeauthor{Fedorova2021} (\citeyear{Fedorova2021}). In that case, during the verification of FRBs, it turned out that the authors incorrectly determine the S/N of the events being verified, and as we expect, incorrectly subtract the baseline. From our perspective, a very thorough  verification of all published events is necessary.

\section*{Acknowledgements}
The study was carried out at the expense of a grant Russian Science Foundation (RSF) 22-12-00236 (https://rscf.ru/project/22-12-00236/).

\end{document}